\documentclass[sigconf]{acmart}

\usepackage{acmart-taps}
\usepackage{rotating}
\usepackage{graphicx}
\usepackage{subfigure}
\usepackage{comment}
\usepackage{listings}
\usepackage{multirow}
\usepackage{pifont}
\usepackage{amsthm}
\usepackage{color}
\usepackage{epstopdf}
\usepackage{enumitem}
 
\let\mathscr\relax
\usepackage{algorithm}
\usepackage{arevmath}     
\usepackage[noend]{algpseudocode}
\usepackage{algpseudocode}
\usepackage{multicol}
\usepackage{balance}
\usepackage{acmart-taps}

\setlist{leftmargin=5.5mm}
\usepackage{tikz}
\usetikzlibrary{shapes.geometric}
\usetikzlibrary{arrows}

\copyrightyear{2023}
\acmYear{2023}
\setcopyright{acmlicensed}
\acmConference[ACSAC '23]{Annual Computer Security Applications Conference}{December 4--8, 2023}{Austin, TX, USA}
\acmBooktitle{Annual Computer Security Applications Conference (ACSAC '23), December 4--8, 2023, Austin, TX, USA}
\acmPrice{15.00}
\acmDOI{10.1145/3627106.3627132}
\acmISBN{979-8-4007-0886-2/23/12}

\usepackage{xspace}
\usepackage{anyfontsize}

\definecolor{dkgreen}{rgb}{0,0.6,0}
\definecolor{gray}{rgb}{0.5,0.5,0.5}
\definecolor{mauve}{rgb}{0.58,0,0.82}

\newcommand{\ttt}[1]{\texttt{#1}}
\newcommand{\paratitle}[1]{\noindent{\em \bf #1}}

\newcommand{\securedl}{{\scshape SecureDL}\xspace}

\newcommand{\containers}{\alpha}
\newcommand{\containerram}{\beta}
\newcommand{\nodecores}{\gamma}
\newcommand{\executorcores}{\delta}
\newcommand{\nodeexecutors}{\epsilon}
\newcommand{\workers}{\omega}

\newcommand{\evalTotalRepos}[0]{2120}
\newcommand{\evalKeywords}[0]{`spark example', `spark tutorial', `spark learning', etc.}
\newcommand{\evalTotalMaven}[0]{637\xspace}
\newcommand{\evalTotalBuiltSuccess}[0]{417\xspace}
\newcommand{\evalTotalBuiltSuccessPercent}[0]{65.46\%\xspace}

\newcommand{\evalTotalAnalyzableJars}[0]{247\xspace}

\newcommand{\evalJarWithIssues}[0]{21\xspace}

\newcommand{\evalRuleOrgApacheSparkPackage}[0]{12\xspace}
\newcommand{\evalRuleInvokingRestrictedClasseForName}[0]{7\xspace}
\newcommand{\evalRuleInvokingRestrictedNetwork}[0]{8\xspace}

\lstdefinelanguage{myScalastyle}{
  keywords={typeof, new, true, def, false, try, catch, function, return, null, catch, switch, var, val, if, in, while, do, else, case, break,let, const, throw},
  language=scala,
  aboveskip=3mm,
  belowskip=3mm,
  showstringspaces=false,
  columns=flexible,
  basicstyle={\small\ttfamily},
  numbers=none,
  numberstyle=\tiny\color{gray},
  keywordstyle=\color{blue},
  ndkeywordstyle=\color{blue}\bfseries,
  commentstyle=\color{dkgreen},
  stringstyle=\color{mauve},
  breaklines=true,
  breakatwhitespace=true,
  tabsize=3,
}

\begin{document}

\title{The Queen's Guard: A Secure Enforcement of Fine-grained Access Control In Distributed Data Analytics Platforms}

\author{Fahad Shaon}
\authornote{These authors contributed equally to this work}
\email{fahad@datasectech.com}
\affiliation{%
  \institution{Data Security Technologies}
  \city{Dallas}
  \state{Texas}
  \country{USA}
}

\author{Sazzadur Rahaman}
\authornotemark[1]
\email{sazz@cs.arizona.edu}
\affiliation{%
  \institution{University of Arizona}
  \city{Tucson}
  \state{Arizona}
  \country{USA}
}

\author{Murat Kantarcioglu}
\email{murat@datasectech.com}
\affiliation{%
  \institution{Data Security Technologies}
  \city{Dallas}
  \state{Texas}
  \country{USA}
}

\begin{abstract}
Distributed data analytics platforms (i.e., Apache Spark, Hadoop) provide high-level APIs to programmatically write analytics tasks that are run distributedly in multiple computing nodes. The design of these frameworks was primarily motivated by performance and usability. Thus, the security takes a back seat. Consequently, they do not inherently support fine-grained access control or offer any plugin mechanism to enable it, making them \textit{risky} to be used in multi-tier organizational settings.

There have been attempts to build ``add-on'' solutions to enable fine-grained access control for distributed data analytics platforms. In this paper, first, we show that straightforward enforcement of ``add-on'' access control is insecure under \textit{adversarial} code execution. \textit{Specifically, we show that an attacker can abuse platform-provided APIs to evade access controls without leaving any traces.} Second, 
we designed a two-layered (i.e., \textit{proactive} and \textit{reactive}) defense system to protect against API abuses. On submission of a user code, our proactive security layer statically screens it to find potential attack signatures prior to its execution. The reactive security layer employs code instrumentation-based runtime checks and sandboxed execution to throttle any exploits at runtime. Next, we propose a new fine-grained access control framework with an enhanced policy language that supports \textit{map} and \textit{filter} primitives. Finally, we build a system named {\bfseries \scshape SecureDL} with our new access control framework and defense system on top of Apache Spark, which ensures secure access control policy enforcement under adversaries capable of executing code. To the best of our knowledge, this is the first fine-grained attribute-based access control framework for distributed data analytics platforms that is secure against platform API abuse attacks. Performance evaluation showed that the overhead due to added security is low.
\end{abstract}

  

\maketitle

\section{Introduction}
In recent years, the capability of collecting information and its usage is increasing at an exponential rate. Consequently, the big data market is also growing significantly~\cite{big_data_growth_forbes}. To process this exorbitant amount of data~\cite{DBLP:journals/monet/ChenML14}, one of the most popular approaches is to use distributed data processing frameworks~\cite{DBLP:conf/osdi/DeanG04, DBLP:conf/nsdi/ZahariaCDDMMFSS12, DBLP:conf/sosp/ZahariaDLHSS13, DBLP:conf/sigmod/ArmbrustXLHLBMK15, DBLP:journals/cacm/ZahariaXWDADMRV16, DBLP:journals/jmlr/MengBYSVLFTAOXX16}, such as Apache Spark~\cite{spark}, Hadoop~\cite{hadoop}, Hive~\cite{hive}, and Pig~\cite{apache_pig}, which can be scaled to process an increasing amount of data by adding more computing nodes. 
Typically, these systems are protected with basic access control and security mechanisms. Most use the access control models provided by the underlying distributed file system protections~\cite{DBLP:conf/osdi/DeanG04, DBLP:conf/nsdi/ZahariaCDDMMFSS12, DBLP:conf/sosp/ZahariaDLHSS13}, enforced only at the file level. Such coarse-grained access control provided by the default is insufficient for many applications. Hence, this limitation spurred further research on enabling \textit{fine-grained} access control for these distributed data analytics platforms.

One approach to enable fine-grained access control protections is by providing higher-level abstractions such as SQL (e.g., Hive built on MapReduce and HDFS, or DeltaLake built on top of Apache Spark). In these approaches, since user-submitted code is restricted to SQL and the underlying data is relational, existing access control solutions developed for relational databases become directly applicable. However, according to multiple sources~\cite{mongo_unstructured, mit_unstructured}, a vast majority (80\% to 90\%) of data produced these days are unstructured, which does not fit into the relational model. 

To support unstructured data processing, these frameworks allow users to submit code written using the framework-provided APIs. 
However, these frameworks' lack of plugin support makes it harder to implement a fine-grained access control mechanism for user-submitted jobs with arbitrary code. One approach to sidestep this limitation is to use ``add-on'' solutions like code instrumentation with inline reference monitors (IRM)~\cite{DBLP:conf/bigdata/UlusoyKPH14, DBLP:conf/ccs/UlusoyCFKP15}. IRM is a technique to enforce security policies by injecting security checks into an untrusted code before execution. For example, in~\cite{DBLP:conf/ccs/UlusoyCFKP15}, authors proposed GuardMR to enable \textit{fine-grained} access control by instrumenting user-submitted jobs to enforce access policies in Hadoop. This and similar instrumentation-based access control enforcement frameworks usually assume that such instrumentation is secure just with the support of the underlying Java virtual machine (JVM) security policies.

In this work, we show that platform-provided sandboxing (i.e., security managers in JVM) alone is inadequate to prevent all the attack surfaces. For example, Java security managers only protect access modification of already \textit{access-protected} methods or fields. It doesn't guard against invoking \textit{public} methods, which an attacker can leverage to evade IRM-based security enforcement. We are the first to systematically study and show that an attacker can abuse platform-provided APIs to evade access controls without leaving any traces, which we call \textit{transient attacks}. Since transient attacks are stealthy by design, defending against them is important to ensure a secure operation of these ``add-on'' solutions, which motivates the following research question: \textit{Is it possible to securely enforce ``add-on'' fine-grained access control policies on the user-submitted code by throttling transient attacks?}.

To address this research question, next, we systematically analyze the causes behind transient evasion attacks and design a two-layered (i.e., \emph{proactive} and \emph{reactive}) defense platform, to protect against them with minimal usability and performance overhead.
Proactive security is enforced before a user's request reaches the data framework. On the other hand, reactive security is enforced inside or alongside the data framework. In the proactive part, we utilize state-of-the-art static program analysis to detect potentially malicious user code that can be used to evade fine-grained access control enforcement. More specifically, we use static program analysis to screen users' code against some predefined rules.
However, some of these rules do not guarantee soundness, which is fundamentally limited by the capability of static analysis-based approaches. We use reactive defense as a \textit{safety-net} for them. Our reactive defense consists of binary integrity checking, static code instrumentation-based runtime checking, and Java security manager. Our binary integrity checking phase ensures the integrity of our trusted computing base (TCB), i.e., system-specific jars. Runtime checks guard against cases when an attacker can bypass the proactive defense to use an adversarial coding capability. Although, because of incurring runtime overheads, we avoided security managers as much as possible; however, it was not feasible to avoid it altogether (Details in Section~\ref{sec:reactive_defence}). 

Next, we propose a new fine-grained attribute-based access control framework that uses Scala as the policy specification language to support enforcing versatile policies on unstructured data. Although there have been some efforts to define access control models for big data analytics systems (e.g., Vigiles~\cite{DBLP:conf/bigdata/UlusoyKPH14}, GuardMR~\cite{DBLP:conf/ccs/UlusoyCFKP15}, compared to previous work, our access control offers the following novel contributions: \textit{i)} since it modifies the user’s submitted code to attach the access control logic to it by leveraging aspect-oriented programming, it is framework-agnostic, \textit{ii)} it supports both \textit{map} and \textit{filter} primitives to support versatile policies for filtering and obfuscating (masking) data. 
We implemented our access control with the two-layered defense on Apache Spark and named it {\bfseries \scshape SecureDL}.
We used aspect-oriented programming to implement the access control due to the lack of a built-in fine-grained access-control plugin system in Apache Spark. We leveraged the two-layered defense to ensure secure policy enforcement under API abuse attacks, as Apache Spark supports arbitrary code execution. In addition, we implement the proposed access control system on Hive to show the framework-agnostic nature of it using its built-in plugin system. \\

Our contribution can be summarized as follows:

\begin{itemize}

    \item We are the first to systematically show that it is possible to evade ``add-on'' solutions that leverage code instrumentation with inline reference monitors to implement fine-grained access controls without leaving any evasion traces, which we call \textit{transient evasion attacks}.
    
    \item We propose a two-layered, proactive, and reactive defense mechanism to protect against these attacks. We show that a combination of static program analysis, sandboxing, and runtime checks can be used to provide protection with low-performance overhead.
    
    \item We provide a new framework-agnostic, fine-grained attribute-based access control mechanism, which supports both \textit{map} and \textit{filter} primitives for versatile policies supports. To demonstrate its wide applicability, we integrated it with frameworks with plugin support (i.e., Hive) and without plugin support (i.e., Spark) for fine-grained access control. We adopted our two-layered defense to secure it under API abuse attacks.

    \item Our experimental evaluation showed that our proposed proactive and reactive defense is effective and incurs low-performance overhead. In a \ttt{6}-node Hadoop cluster, we observe only about \ttt{4\%} overhead on average on processing queries from TPCH benchmark.
\end{itemize}

\section{Background and threat model}
\label{background}

In this section, we first provide a background on Apache Hadoop and Spark that would be useful for understanding our attacks. Then, we briefly discuss the threat model. 

\subsection{Background}

\noindent
{\textbf{Apache Hadoop architecture.}} Apache Hadoop consists of two significant components -- a distributed file storage system (HDFS) and a job execution framework (YARN). HDFS splits files into multiple parts and saves them on different machines. It also replicates splits for high availability. YARN executes user-submitted map-reduce code in a distributed manner. 

A job in the MapReduce system has a few key components. The user defines the input data format by implementing - InputSplit with details of data chunks, RecordReader containing details of how to read records from these chunks, and InputFormat containing details of how to create InputSplit and RecordReader objects. The user also defines the computation with map-reduce functions, where Hadoop executes the map method once for each record. The map job emits key-value pairs, which Hadoop combines according to the keys and invokes the reduce method once per key.

\noindent
\textbf{Apache Spark architecture.}
Like other data processing frameworks, Apache Spark also utilizes the distributed data processing paradigm, where there is a master node (known as \textit{driver}) that receives a data-analytic task and distributes it to various other workers nodes (known as \textit{executors}). In Spark, a user can submit jobs written in a Turing complete language, and the system executes the code in a distributed manner.
Typically, the Apache Spark cluster operates in two modes, i.e., i) standalone and ii) interactive. 
In a standalone mode, a user can submit a job jar to a spark cluster via \textit{SparkSubmit} shell. The driver node accepts the submitted jar and creates a \texttt{SparkContext} within itself, which prepares and sends specific tasks to the executors.

In the interactive mode, users submit code from interactive notebooks (i.e., Zeppelin, Jupyter, etc.), which use Livy for interactive job execution in a Spark cluster. Livy is an open-source REST interface for interacting with Spark. Livy acts as a driver in this setting and supports executing code snippets or an entire program. While running from multiple users, Livy relies on user emulations, such as user proxy, to emulate their capabilities.

\lstset{xleftmargin=-4mm}

\begin{lstlisting}[caption={An example use of Spark RDDs. Here, arrows represent the pointer from a child RDD to its parent RDD.},label={rdd:example},escapeinside=||]
  |\includegraphics[width=0.95\linewidth]{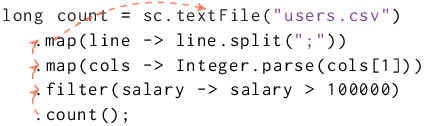}|
\end{lstlisting}

\lstset{xleftmargin=4mm}

\paratitle{RDD, DataSet, and DataFrame.} Resilient distributed dataset (RDD) is a fundamental data structure in Apache Spark that abstracts a collection of elements residing across the nodes in a Spark cluster and supports a predefined set of operations on it in a fault-tolerant manner. RDD operations are of two types: i) transformations and ii) actions. Transformations create a new RDD from an existing one, and the actions return a value to the driver program after running a computation on the transformed dataset (if a transformation is applied). Typically, initial RDDs are created from files persisted in a distributed file system (e.g., HDFS). Listing~\ref{rdd:example} presents an example of Spark RDD. Given a file, \textit{users.csv} with users and their salaries, the goal is to find the number of users with a salary of at least 100K. Here, \textit{textFile} creates the \textit{initial RDD}, \textit{map}, and \textit{filter} are two transformations. Given an RDD, both \textit{map} and \textit{filter} return a new instance of an RDD after applying the transformation defined in the argument. \textit{count} is an action that returns the count of the elements of a given RDD. Spark remembers the transformations by creating a directed acyclic graph (DAG) of all operations. Arrow in Listing~\ref{rdd:example} represents the parent-child relationship among RDDs in their DAG representation. Similar to RDD, both DataSet and DataFrames are immutable collections of distributed and partitioned data~\cite{data_frame_spark}.

\subsection{Threat model}
\label{subsec:threatmodel}

\noindent
\textbf{Attacker Goal:} Our attacks aim to evade fine-grained access control transiently (i.e., without leaving any traces) on distributed data analytics frameworks by abusing the platform-provided APIs.

\noindent
\textbf{Assumptions:} \textit{We also assume that \underline{our attacker is an insider}, who is a data analytic user with lower access privilege \textit{i)} can run code for data analytics tasks and \textit{ii)} has incentives to evade such access control if the chance of getting caught is low.} We consider data lakes in a multitier organizational setting, where data access is managed by distributed data processing engines. The access is controlled on a need-to-know basis with a fine-grained access control mechanism. We assume the fine-grained access controls are implemented by leveraging code instrumentation with inline reference monitors (IRM). We consider framework-specific runtime and jars as our trusted computing base (TCB), which means these components are trusted, and any tempering of them is detectable. \textit{Why TCB?} TCB provides a trust anchor for our defense. Any tempering of this would limit the guarantees of the proposed defense (Section~\ref{sec:defending_securedl}).

\noindent
\textbf{Attack severity:} Insider attacks are a real threat in a multi-tier organization. This is one of the leading causes of major fraud in the telecom~\cite{telcom} and financial sector~\cite{fintech}. In theory, the threat model assumed in this paper would enable attackers to access sensitive information by evading access controls, and that too without leaving any traces. Additionally, this type of attack could apply to future systems that may want to support the NIST ABAC~\cite{niststd} standard where complex evaluations are needed.

\section{Attacks on IRM-based Approaches}
\label{sec:attacking:irms}

Inline reference monitoring (IRM) allows a convenient way to inject security enforcement into a system. In JVM-based frameworks, Aspect Oriented Programming~\cite{kiczales1997aspect_aop} is typically used to implement IRMs~\cite{DBLP:conf/ccs/UlusoyCFKP15, DBLP:conf/bigdata/UlusoyKPH14}. For example,
GuardMR injects access control logic (known as \emph{advice}) to existing target functions or framework hooks (known as \emph{pointcut}) without changing the source code at runtime. Thus, to evade such enforcement, it is sufficient to find ways to access data by avoiding the IRM hooks. We leverage this insight to craft concrete attacks against IRM-based access control enhancements~\cite{DBLP:conf/ccs/UlusoyCFKP15, DBLP:conf/bigdata/UlusoyKPH14} atop distributed data analytics platforms. Before we present our attacks, we first provide the details of the IRM implementation for both Hadoop and Spark.

\subsection{Attacking IRMs on Hadoop}

Vigiles~\cite{DBLP:conf/bigdata/UlusoyKPH14} and GuardMR~\cite{DBLP:conf/ccs/UlusoyCFKP15} used IRM-based approach to implement fine-grained access control for Hadoop. Since GuardMR~\cite{DBLP:conf/ccs/UlusoyCFKP15} is the most recent work, we will use GuardMR's implementation to demonstrate our attack, which can be trivially extended for Vigiles.

In GuardMR, when users submit jobs with InputFormat, InputSplit, and RecordReader definitions, it constructs a new InputFormat, InputSplit, and RecordReaders that wraps these methods with policy enforcement. GuardMR uses aspect-oriented programming to implement IRM to detect target methods and inject access control policies into them. Let the provided untrusted function (e.g., a user-provided RecordReader) be $f_i$, GuardMR builds a new method $f_o$ such that $f_o = f_i(f_e)$. Here, $f_e$ reads the original data and applies relevant policies, i.e., filters and masks the data, then forwards the data to function $f_i$.

\noindent
{\textbf{Scenario \#1: Reading with RecordReader.}} To attack GuardMR, an attacker has to figure out the original data. It boils down to figuring out the files/splits to access by using the Hadoop-provided APIs. It turns out that, inside a provided custom \texttt{RecordReader}, the attacker can easily read the original input stream and access the data directly, which sidesteps the policies enforced in $f_e$, which GuardMR injects. Note that such features are not preventable with sandboxing alone without hurting legitimate functionalities. The attack code snippet is presented in Appendix~\ref{sec:attack:guradmr}.

\subsection{Attacking IRMs on Spark}
\label{sec:attack:irm:spark}

\noindent
\textbf{IRMs on Spark.} Currently, no solutions exist to implement fine-grained access controls on Apache Spark allowing arbitrary code execution. execution. Hypothetically, the most convenient places to implement such solutions on Apache Spark are RDD or DataFrame creation methods, such as \texttt{org.apache.spark.SparkContext.textFile(...)}, \texttt{org.apache.spark.sql.DataFrameReader.json(String)}, etc. 
Because Spark data users are restricted only to use these methods to create an initial RDD and perform various operations.Like GuardMR, one can inject specialized transformations (i.e., map, filter) to enforce access control on the initial RDD and then return it so that all the user-defined operations are executed after the policy enforcement. However, a user bypassing the execution of the specialized transformation by retrieving the initial RDD will be able to evade the access control enforcement. Interestingly, each RDD contains an internal reference to its parent RDD (Listing~\ref{rdd:example}) and the initial RDD. If an attacker can access these references, they would be able to retrieve the initial RDD before adding new operations. The following attacks will leverage this fact in two different ways.

\begin{lstlisting}[caption={Retrieving the reference to the initial RDD with Java Reflection to bypass IRM-based access control.},label={rdd:reflection}]

val rd = sc.textFile("users.csv")
val clazz = rd.getClass

// #1. Read with "prev" field
val fld = clazz.getDeclaredField("prev")
fld.setAccessible(true)
val parent = fld.get(rd)
val initParent = fld.get(parent)

// #2. Read with "prev" method
val method = clazz.getMethod("prev")
val parent = method.invoke(rd)
val initParent = method.invoke(parent)

// #3. Read  with "parent" method
val mthd = clazz.getMethod("parent", 0)
val initParent = mthd.invoke(rd, ...)

// #4. Read with "firstParent" method
val method = clazz.getMethod("firstParent")
val initParent = method.invoke(rd, ...)
\end{lstlisting}

\noindent
{\textbf{Scenario \#1: Java reflections.}} To obtain the private properties of an object, an attacker can use reflections. Listing~\ref{rdd:reflection} shows a demonstration of retrieving the initial RDD by accessing a private field \textit{prev} of the RDD, which contains the reference to its parent. The \textit{prev} field also has a corresponding package-private method named \textit{prev} to provide easy access within the spark framework codes, which can also be used similarly. Some RDDs also have a package-private method \textit{parent}, which can be used to access any parents, and a convenient method \textit{firstParent}, which directly returns the reference to the initial RDD.

\begin{lstlisting}[caption={Retrieving the reference to the initial RDD with spark specific package naming},label={rdd:package}]
val rd = sc.textFile("users.csv")
// accessing the parent pointer
// with "parent" method
val parent = rdd.parent(0)
\end{lstlisting}

\noindent
{\textbf{Scenario \#2: Spark-specific package.}}
If a user defines a class in a package named ``org.spark.*'', builds the jar and puts it into the classpath. While in execution, there is no distinction between the user's package and those from Apache Spark. As a result, a class in the user's package can access all package-private methods and fields without even requiring reflection.
\textit{firstParent} and \textit{prev} Methods are package-private, which means these methods are accessible within ``org.spark.*''. However, an attacker can create their class with the same prefix to directly invoke the methods from a Spark job (Listing~\ref{rdd:package}).

\section{Defense Methodology}
\label{sec:defending_securedl}

Section~\ref{sec:attacking:irms} shows that the attacks' nature mainly depends on the framework APIs used for AOP hooks, and the usage and functionality of these APIs vary across different frameworks. Since designing a general defense covering all the frameworks is challenging, we focus on Apache Spark in this paper. Apache Spark represents the group of frameworks (i.e., Spark, Hadoop, Flink, etc.) that support code execution.

\noindent
\textbf{Defense goal.} Given an Apache Spark task $J$ with arbitrary code, detecting if it is malicious is undecidable, in general~\cite{DBLP:journals/compsec/Cohen87}. To avoid this pitfall, instead of detection, our goal is to prevent transient evasion with minimal performance overhead without hurting legitimate uses. Toward that goal, we will first systematically analyze the platform APIs that offer the primary attack surface and show how those surfaces can be nullified. \textit{Note that this is the first stab towards protecting distributed data analytic systems from a powerful adversary like this, which might stimulate future research.}

\subsection{Attack surfaces in Spark}
\label{sub:adversarial_coding_capabilities}

In this section, we discuss the APIs of Apache Spark that might offer attack surfaces for transient evasion. To securely deploy and maintain secure operations of IRM-based access control, these APIs need to be restricted. 

\begin{enumerate}
 
\item \textbf{Restricting reflection on RDDs.}
Java reflection API allows users to access an object's private properties (fields and methods). Specifically, an attacker can use Java reflection APIs (attack \#3) to bypass the \securedl access control protection. An intuitive approach to protect against reflection is to sandbox the spark job execution with Java security manager~\cite{DBLP:conf/ccs/UlusoyCFKP15}. However, security managers can only protect against access modification and retrieving declared methods or fields. However, it does not guard against invoking \textit{public} methods. 

When Spark's internal Scala classes are compiled into Java class files, all the package-private methods become public. Because of this, it does not require performing any access modification while invoking any of the \textit{prev}, \textit{parent}, \textit{firstParent} methods in Listing~\ref{rdd:reflection}. Thus, a security-manager-based solution is insufficient, and a more robust sandboxing mechanism is required to prevent this attack.

\item \textbf{Preventing framework-specific package declarations.}
In attack \#4, we see that an attacker can define ``org.spark.*'' package to directly invoke \textit{prev}, \textit{parent}, \textit{firstParent} methods in scala. Spark jobs must be vetted against such manipulations.

\item \textbf{Preventing dynamic class loading.} Java allows users to load a class dynamically, given a class name. This allows the user to load any class in the current classpath. This capability can potentially enable an attacker to execute non-screened codes, including code instrumentation.

\item \textbf{Preventing to override security managers.} A security manager is a class that defines the security policies of an application. It has an implementation of several check* methods, such as checkPermission, checkWrite, and checkExec. These methods determine whether particular actions, such as writing a file, are permitted in the current running Java virtual machine instance. Security managers are typically used to build a sandboxed/protected execution environment. Interestingly, a user can replace an existing security manager with code if it is not configured correctly. This replacement mechanism can be leveraged to bypass the existing protections. To build a secure system, replacing existing security managers and setting up custom policies must be disabled.

\item \textbf{Preventing native codes and libraries.} 
One can use native codes or libraries to enable any of the features restricted in the Java layer. Hence, loading native libraries and performing native API calls must be flagged.

\end{enumerate}

Note that, we consider the APIs for system command execution and file/read writes to be out of the scope, since, attacks leveraging these APIs for out-of-the-ordinary operations would leave strong signatures for detection~\cite{DBLP:conf/uss/DattaPI0E22}. Thus it might not be in the best interest of the attackers who do not wish to leave traces to use these APIs.

\subsection{Defense overview}\label{sec:defense}
In this section, we propose a combination of proactive and reactive mechanisms to restrict the adversarial coding capabilities by leveraging Spark APIs discussed in Section~\ref{sub:adversarial_coding_capabilities} to ensure an automated secure operation. In theory, we can block all the executions of problematic APIs and trivially protect the system. However, the question that needs an answer is: \textit{``Is it possible to guarantee a secure prevention of adversarial capabilities with minimal overheads without denying services to legitimate users?''}

In this section, we answer this question affirmatively. Intuitively, the overhead would be minimal if all adversarial capabilities could be prevented proactively by analyzing the submitted code before execution. \textit{Therefore, we use static analysis to prevent most of the attack surfaces (or APIs) and avoid costly runtime checks (i.e., security managers) as much as possible.} In our design we first categorize the attack surfaces into two groups, \textit{i)} blockable attack surfaces, and \textit{ii)} non-blockable attack surfaces. An attack surface is blockable if all of its instances can be blocked since this does not appear in typical usage scenarios. Simple static analysis methods can be designed to proactively detect and block them with \textit{soundness} (no miss detections) and (almost) \textit{completeness} (no false alarms) guarantees. 
However, not all attack surfaces are of this nature. For example, Apache Spark uses reflection APIs in its regular operations, thus blocking it altogether infeasible. We call such attack surfaces non-blockable. For non-blockable attack surfaces (APIs), we design static analysis methods to identify their \textit{malicious usage} for proactive detection. However, existing static analysis techniques for API misuse detection are known to be unsound (missed detection) and incomplete (generates false alarms)~\cite{rahaman2019cryptoguard, DBLP:conf/ecoop/KrugerS0BM18}. 
\textit{Missed detections} will enable attackers to evade the defense and \textit{false alarms} will deny services to benign users. Thus, our goal to protect non-blockable attack surfaces proactively, we use \textit{``soundy''} (means mostly sound~\cite{DBLP:conf/uss/MachirySCSKV17}, but no guarantees) dataflow analysis framework with fewer false alerts.
\textit{For this case, we also employ a reactive fall-back mechanism to block adversarial uses that evade proactive defense.} The goals of the reactive mechanism are as follows, \textit{i)} ensure a sandboxed execution of the user-submitted code, and \textit{ii)} block abuse of non-blockable APIs that evaded proactive safety checks.

\begin{figure*}[!ht]
    \begin{center}
    \scriptsize
    \begin{multicols}{2}
     \centering
     \begin{equation*}
    \text{~~~~~~~~~~~~~~~~~~~~~~~~~~~} Block\cfrac{{
     {\color{red}Get}\cfrac{{p_{o} : \textit{{\color{red}get}(\textbf{obj})}, ~x : x~ \textbf{is an}~ RDD, ~V : \{v_{i}\} ~|~ v_{i} \rightsquigarrow p_{o}, \forall i \in \text{[}1, |V| \text{]}}}{if~x~\epsilon ~V ~then ~\textbf{true}~ else~ \textbf{false}},~~
     {\color{red}Invoke}\cfrac{{p_{o} : \textit{{\color{red}invoke}(\textbf{obj}, \_)},~ x : x~ \textbf{is an}~ RDD, ~V : \{v_{i}\} ~|~ v_{i} \rightsquigarrow p_{o}, \forall i \in \text{[}1, |V| \text{]}}}{if~x~\epsilon ~V ~then ~\textbf{true}~ else~ \textbf{false}}
     }}{{if~{\color{red}Get} ~\textbf{or} ~{\color{red}Invoke} ~then ~\textbf{true}~ else~ \textbf{false}}}
    \end{equation*}
    \end{multicols}
        \vspace{-15pt}
    \caption{Blocking the use of reflection on RDD objects. Here,  $v_{i} \rightsquigarrow p_{o}$ represents an influence of an object $v_{i}$ on the program point $p_{o}$. $V$ represents the set of all such objects.}
    \label{fig:backward}
    \end{center}
    \vspace{-4mm}
\end{figure*}

\subsection{Proactive defense}

In this section, we present the proactive agent, which uses static code analysis to screen the user-submitted code. Screening is done by checking the code against some well-defined rules. We have two types of rules, \textit{i)} rules for blockable attack surfaces, and \textit{ii)} rules for non-blockable attack surfaces, which we discuss next.

\subsubsection{\textbf{Restricting blockable attack surfaces}}
If an attack surface does not often appear in regular use cases, we prescribe blocking that as a whole, which we refer to as the blockable attack surface. We use regular expressions to implement these rules for sound and (almost) complete detection.

\underline{\textit{Restricting framework-specific packages.}} For security purposes, Apache Spark intentionally put some of the framework's internal APIs as \textit{package-private}, so that these APIs are hidden from external users. As we discussed in Section~\ref{sub:adversarial_coding_capabilities}, a user can define classes with the framework-specific package structure with a prefix of ``\textit{org.apache.spark}'', so that the framework internal APIs become accessible. 
Therefore, to prevent the adversarial use of these capabilities, we block jobs that leverage this capability to invoke the APIs to access the parent objects of an RDD (e.g., ``prev'', ``parent'') as demonstrated in Listing~\ref{rdd:package}. 
        
\underline{\textit{Restricting permissive system APIs.}} We also restrict users to invoking the following system APIs (1) to load classes dynamically; (2) to override the security manager; (3) using native codes/libraries, etc. One might argue that an attacker could use third-party libraries that are not covered by our defense. However, in such cases, the attacker is also needed to embed the libraries within her submitted, which would use the standard APIs we cover.

In a real environment, there might be instances where blocking specific instances of the above cases is infeasible. We design a allowlisting mechanism (Section~\ref{extend:tcb}) to handle these cases.

\subsubsection{\bf{Restricting non-blockable attack surfaces statically.}}\label{restrict:non:blockable} It is infeasible to block all the permissive system APIs as a whole. For example, several machine learning libraries benignly use reflection APIs for optimizing job performance. Therefore, to prevent the adversarial use of these capabilities, it is infeasible to block all of their usage. We handle the reflection APIs as follows. 

\begin{enumerate}
    \item We use security managers to block unusual cases of reflection APIs (Section~\ref{security:manager}). This is because security manager-based sandboxing is sound by design.

    \item However, Java security managers are inadequate to guard against invoking public methods, which an attacker can leverage to evade IRM-based security enforcement (Attacks 2, 3, 4 in Listing~\ref{rdd:reflection}). If it is infeasible to use security managers, we leverage existing advances in static dataflow analysis-based API abuse detection~\cite{rahaman2019cryptoguard, DBLP:conf/ecoop/KrugerS0BM18} to proactively detect malicious use of these APIs. To minimize the impact of false positives for most common usage, we leverage a allowlisting mechanism discussed in Section~\ref{extend:tcb}. 
    
    \item If it is infeasible to model the abuses with dataflow analysis, we allowlist (Section~\ref{extend:tcb}) all the common uses of the API and block all the others that are not allowlisted.
\end{enumerate}
 
Note that, the static dataflow analysis does not guarantee soundness. We rewrite the user-submitted Spark jobs to ensure runtime checks for the cases that are missed during our proactive phase. If misuse is detected, our runtime fallback mechanism blocks further execution, which guarantees the security of the framework (Section~\ref{para:runtime_check}). Next, we show an example of designing static dataflow analysis to detect malicious uses of a system API in the light of reflection APIs.

\underline{\textit{Detection of reflection API abuses.}} In Section~\ref{sub:adversarial_coding_capabilities}, we observe that to obtain the private properties of an object, we need to invoke \textit{java.lang.Object get(java.lang.Object)} on the corresponding field by using the object of interest as the parameter. Similarly, to invoke methods on an object, it is required to invoke \textit{java.lang.Object invoke(java.lang.Object,java.lang.Object[])} on the corresponding method by using the object of interest as the first parameter. Java security managers cannot sandbox the execution of the ``get'' and ``invoke'' methods on public properties of a  class (cases 2, 3, 4 in Listing~\ref{rdd:reflection}).
Thus, we use backward data-flow analysis to detect and block jobs that leverage these APIs to access the parent objects of an RDD. Specifically, our backward data-flow analysis identifies whether an RDD instance passes as an input parameter to these methods, which is formally defined in Figure~\ref{fig:backward}.

\textbf{Backward dataflow analysis implementation.} 
Since implementing a new dataflow analysis is not our main focus, we use the interprocedural backward data-flow implementation of CryptoGuard~\cite{rahaman2019cryptoguard} for this purpose. CryptoGuard's implementation is demand-driven and known to produce fewer false alarms, which is suitable for our case. However, like all the other static dataflow analysis frameworks, it does not guarantee soundness. \textit{Note that to improve the performance (i.e., improve robustness or runtime performance, reduce false positives or false negatives, etc.), CryptoGuard can be replaced with any other competing solutions~\cite{DBLP:conf/ecoop/KrugerS0BM18, DBLP:journals/pacmpl/SpathAB17}, which is beyond the scope of this work.}

\subsubsection{\bf{Extending the trusted computing base (TCB)}}\label{extend:tcb}
Identifying the code that is controlled by a malicious user is instrumental to provide seamless service to the legitimate users. For example, various third-party libraries use Java Reflection APIs to offer convenient utility. If the code analysis engine wrongfully rejects a job with Java Reflection API invocation, a legitimate user using such libraries will be impacted. To solve this problem, we offer library \textit{allowlisting service}. Our code analysis engine will skip the screening of a jar or class binary if it is allowlisted. We created a list of common libraries that are allowlisted by default, which is considered to be part of our trusted computing base (TCB). The list can be extended or modified by an administrator. 
To allowlist a jar, first, we compute the hash of the jar and store it. Then we unzip it and compute the hashes of each of the class files (or native codes) and store them in a database. 
During the static analysis of a jar, the analyzer first creates a hash of the jar and looks up the database to see whether it exists in the allowlist or not. If found, then the analysis engine skips it. Otherwise, it unzips the jar and creates hashes for each of the class files from inside the jar. If the hash of a corresponding class is not found, then the class is included in the static analysis, otherwise, it is skipped. 
To improve the performance, our analysis engine can also maintain a cache of the analyzed code. If the analysis result of a jar or a class is available in the analysis cache, then we could retrieve the analysis result from the cache and skip the reanalysis. 

\subsection{Reactive defense}
\label{sec:reactive_defence}
The goal of our reactive defense is two folds, \textit{i)} ensuring the integrity of our trusted computing base (TCB) and \textit{ii)} restrict non-blockable attack surfaces that either escaped or not covered by the proactive defense.
The reactive defense has two types of components - \textit{i)} a static component, that matches the cryptographic hashes of the TCB before running a job, and ii) a dynamic component, which works as a fallback for the non-blockable attack surfaces. The dynamic component consists of two parts (1) Java security manager-based sandboxing of API usage to restrict abuse cases, and (2) rewriting the user-submitted job with runtime checks for the APIs that are not sandboxed.

\subsubsection{\bf{Restricting APIs with security managers}}\label{security:manager}
JVM ecosystem offers security managers to secure sandbox untrusted code. Given the context (call trace with invocation parameters) of a system call invocation, security managers can block its execution (by throwing exceptions), if the operation is not permissible. Permission represents access to a system resource. The list of permissions that can be checked by using security managers can be found here~\cite{security_manager}. We use security managers to block the following reflection permission, i) \textit{accessDeclaredMembers} -- querying public, protected, private properties of a class, ii) \textit{suppressAccessChecks} -- accessing public, protected, private properties of a class, and iii) \textit{newProxyInPackage} -- creating proxy instances of a nonpublic interface in a given package. By checking the invocation parameters, we block all these permissions if they are used to access RDD properties. This effectively blocks the attack \#1 in Listing~\ref{rdd:reflection}.
    
\subsubsection{\bf{Defense with instrumentation-based runtime checks.}}
\label{para:runtime_check} As explained in Section~\ref{restrict:non:blockable}, not all reflection API uses can be restricted with security managers. To detect those cases, we designed static-dataflow analysis-based proactive method (Section~\ref{restrict:non:blockable}). However, since dataflow analysis is not ``sound'', it is possible to craft attacks to evade them. Thus, to guard against this, we introduce a runtime check just before the invocation of these APIs. If it is invoked on an instance of RDD or a sub-class of RDD, we generate a runtime exception. 

\section{A new access control Framework with enhanced policy language}
\label{access:control}

Here, we present a new \underline{framework-agnostic} fine-grained access control with enhanced policy language, which can handle both structured and unstructured data.

\noindent
{\bf{Our new access control features.}}
Although there have been some efforts to define access control models for big data analytics systems (e.g.,  Vigiles~\cite{DBLP:conf/bigdata/UlusoyKPH14}, GuardMR~\cite{DBLP:conf/ccs/UlusoyCFKP15}), compared to previous work our access control mechanism\footnote{closely follows NIST ABAC guideline~\cite{niststd}} enables three benefits which were not achieved simultaneously before. These are: \textit{i)} our access control policies are framework agnostic, \textit{ii)} our policy language supports both \textit{map} and \textit{filter} primitives, which enables to write specialized filter and map tasks and \textit{iii)} our policy language allows Scala code to support the enforcement of versatile policies for \textit{map} or \textit{filter} tasks which are specially suited for unstructured data. This means, that by using our filters, one can define arbitrary filtration on arbitrary attributes. However, our map primitives are limited to regular language, which is incapable of expressing any rule requiring memory (or state). Although the capability can be trivially extended, we believe regular language is sufficient enough to capture most real-world scenarios.

\subsection{Access control framework components}
We represent the input data as a dataframe. Dataframe is a structured data storage that stores data in rows and columns. Formally, an input data set $D_{id} = <id \left ( D  \right), C, T>$ consists of an unique identifier (e.g., filename, table name), column definitions, and an ordered set of tuples. 
This abstraction is very powerful and encompasses a wide variety of data types. Intuitively, a dataframe directly maps to a relational table. In addition, we can represent any non-relational data using this abstraction. For example, this abstraction can be used to model a text file where we assume each line is a tuple of a single element and can assume the name of the element is `text'.
In addition, we can model arbitrarily nested Json data, where each attribute of the input Json becomes an element of the tuple.
In reality, a wide variety of popular data analytics systems represent data in this format, such as Spark~\cite{data_frame_spark}, Pandas~\cite{data_frame_pandas}, R\cite{data_frame_r}, etc.
Furthermore, we can represent nontextual, such as images, data into dataframes by keeping a column of binary data (BLOB in a relational database). This simplifies data processing since we can efficiently manage meta-data as well.

\emph{Policy} in our system defined as $P = \langle\mathcal{I}, A, \mathcal{M}, f\rangle$, 
where $I \left(D_{id},u,A)\right)$ is a boolean function for deciding whether a given dataframe $D_{id}$, a user $u$, and set of attributes $A$, the policy is applicable or not, $\mathcal{M}\left(t, u, c\right)$ is set of masking functions, $f \left(t, u, c\right)$ is user provided boolean function for limiting view of the data applied to each tuple $t \in D_{id}$ using user information $u$ and system context information $c$ (e.g., IP address of the request).

A masking or obfuscation function $m$ in our system takes input of a type of data,  modifies it, and then returns same type of data with limited information (i.e. $type\left(a\right) = type\left(m\left(a\right)\right)$). Let $X\left(regex, s\right)$ be a function that takes a regular expression and a string value and returns the indexes of string regular expression matches, $S\left(matches, s, pattern\right)$ be a substitution function that takes the regular expression matches, original string, and a pattern, return the string with replaced pattern in matching location. For example, a regular expression-based US phone number masking function that only returns the last four digits can be expressed as
\[
index = X\left( `\\(?\\d{3}\\)?(-| )\\d{3}-\\d{4}', s\right)
\]
\[
m_{p}\left(s\right) = S\left(index , s, `***-***-dddd' \right)
\]

\begin{algorithm}
\caption{Policy Enforcement}\label{alg:policy}
\begin{algorithmic}[1]
\Ensure $I \left (D_{id},u,A \right)= \mbox{True}$ \label{alg:check}
\Comment{Ensure that policy is applicable}
\Procedure{Policy Enforcement}{$P, u,c,D_{id}$}
\State
\Comment{Apply policy $P$ for a request for data frame $D_{id}$ submitted by user $u$ given the request context $c$}
  \State $D' \gets \emptyset$
  \ForAll{$t \in D_{id} $}  
   \If{$f \left(t,u,c \right)=\mbox{True}$} \label{alg:filter}
    \State $D' \gets D' \cup M \left(t, a, m \right)$ \label{alg:mask}
   \EndIf 
 \EndFor
\State Return $D'$
\EndProcedure
\end{algorithmic}
\end{algorithm}

For efficiency reasons, we define masking functions specific to a column. Let $M \left (t, a, m \right )$ be a column specific masking function that applies masking function $m$ on column $a$ of tuple $t$, i.e. $M \left( t, c, m \right) = m \left(t.a \right)$. Finally, in $M$ we have ordered sets of masking functions potentially for each different column, $\mathcal{M} = \{M_1, M_2, ...\}$

In summary, given a policy $P$,user $u$ and date frame $D_{id}$, first the system checks whether $I\left(D_{id},u,A)\right)$ returns true. For each tuple $t \in D_{id}$, it checks whether $f\left(t, u, c\right)$  returns true (Line~\ref{alg:filter} in Algorithm~\ref{alg:policy}). Then for all $t \in D_{id}: f\left(t, u, c\right)= \mbox{True}$, it adds the masked version of the tuple $t$ 
to the resulting data frame (Line~\ref{alg:mask} in Algorithm~\ref{alg:policy}).
Since our system allows arbitrary scala code for functions $I,f,m$, it can represent any existing role-based (RBAC)~\cite{rbac} and attribute-based access control policies (ABAC)~\cite{abac}.

Since our system allows us to specify ABAC policies using the Scala programming language, any user and data attributes can be combined with programming languages to enforce very sophisticated security and privacy policies. For example, using a custom-defined function defined on images, a policy that can redact human faces automatically can be defined in our system. In other words, a mask function $M$ defined over images can use an ML subroutine to detect the human faces and replace the detected pixels with black ones to redact human faces. We would like to stress that our policies are generic enough to represent any ABAC policies defined on the dataframe abstraction. As we discussed above, this abstraction can represent policies at any granularity for relational, semi-structured, and unstructured data.

\subsection{Implementation using AOP in Spark}
We implement our access control on Apache Spark and name the system as {\sc SecureDL}. Our goal is to keep the enforcement system as transparent as possible from the data user's point of view, i.e., without introducing new APIs. All existing jobs written using current API calls must work in our new system.
To implement the fine-grained access control in this manner, we have two options - (1) we could rewrite the distributed data analytics system with the necessary enforcement codes, and build our version of Spark (i.e., embed the reference monitor inside the system), (2) use an inline reference monitor (IRM) (i.e. we attach our enforcement logic at run-time)~\cite{irm} . 

For our system, we chose the IRM approach because changing and rebuilding existing systems is difficult and time-consuming. Simply, given a policy $P$, user-submitted job $j$, our policy rewriter will rewrite the job $j$ into $j'$ so that the policy is enforced. For a policy $P$, it maps masking operations with a \textit{map} transformation and filter operations with a \textit{filter} transformation.
To implement IRM-based policy enforcement in our system, we choose Aspect-oriented programming (AOP). 
 We defer the discussion of implementation details to Appendix~\ref{sec:spark_aop_details}. In Section~\ref{sec:attacking:irms}, we discussed several concrete attacks that can evade our IRM-based implementation in Spark. We use the proactive and reactive defenses discussed in Section~\ref{sec:defense} to defend against evasion attacks discussed in Section~\ref{sec:attacking:irms}.

\begin{figure}[h]
\centering
\includegraphics[width=\linewidth]{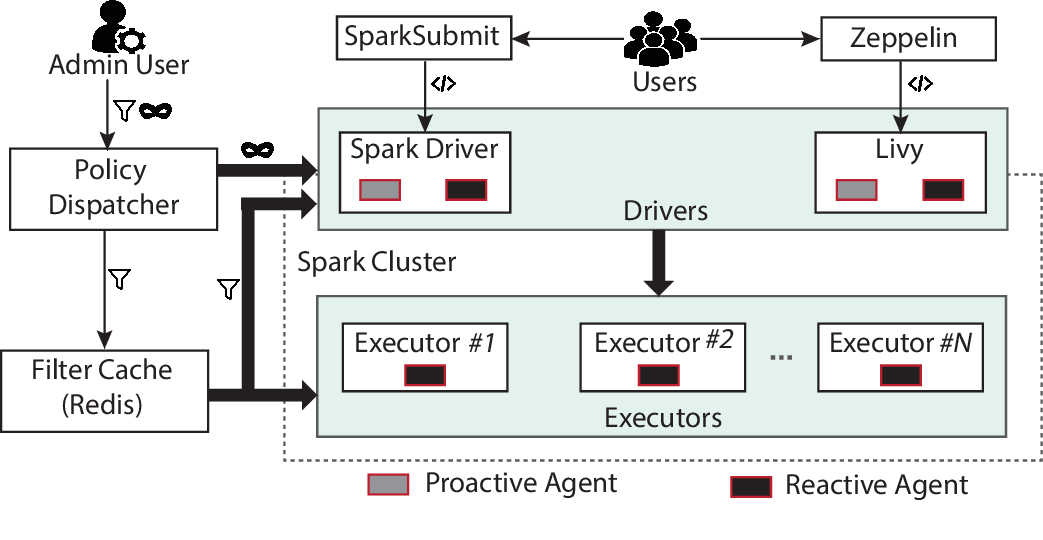}
\vspace{-5pt}
 \caption{System overview of our policy enforcement in Apache Spark with proactive and reactive defenses. Here proactive agents, reactive agents, policy dispatchers, and filter caching are the new components proposed in \securedl.}
 	\label{fig:framework_v2}
\end{figure}

\noindent
\textbf{System overview.}
In this section, we provide an overview of the whole system with defense in place for Apache Spark. 
Figure~\ref{fig:framework_v2} shows the system overview. In this system setup, data analytics users can submit tasks through \textit{SparkSubmit} client and the interactive Zeppelin server \footnote{Apache Zeppelin is a "Web-based notebook that enables data-driven, interactive data analytics and collaborative documents with SQL, Scala and more"~\cite{zeppelin}}. Admin users define attribute-based policies and send them to the policy dispatcher. Policy dispatcher maps the policies into \textit{map} and \textit{reduce} transformations. It bundles all data masking operations into a \textit{map} and arbitrary data filtration logics into a callback method, which can be invoked from a filter. Then, it sends the \textit{map}s into Spark drivers and the callback method binary to a distributed cache implemented with Redis. Along with the map and filter code, this callback method binary is loaded in both drivers and executors. In Figure~\ref{fig:framework_v2}, proactive and reactive agents are employed to guard against bypassing this policy enforcement during data analysis. The proactive agents analyze the submitted code and proactively \textit{rejects a job if detects any bypass attempts}. The reactive agent i) sandboxes the user-submitted code execution by using the  Java security manager and ii) rewrites the user-submitted code by instrumenting runtime checks on certain system API invocations to ensure their secure use. To show the framework-agnostic nature of the proposed access control method, we also present a plugin-based implementation with Apache Hive in Appendix~\ref{hive:implimentation}.

\section{Evaluation}

We performed extensive experiments to quantify the overhead of different components in Apache Spark when our fine-grained access control and defense mechanism is in place. In this section, we present our experimental results.

\paratitle{Cluster configurations.}
We ran experiments on Hadoop Spark clusters with one master node, a few worker nodes, and one service node. 
All these nodes are running inside a virtual cloud network, which is located in a cloud availability zone.
We ran our experiments in Oracle Cloud Infrastructure (OCI) and each node in the cluster is of type \ttt{VM.Standard2.4} having \ttt{4} OCPU, \ttt{60GB} of main memory, running \ttt{Ubuntu 18.04} OS. We also mount a block device disk of size \ttt{1TB} on each instances.
We are using Hadoop version 3.3.0, Spark 3.0.1, and Livy 0.8.0 snapshot (HEAD 4d8a912). Also, our trusted computing base (TCB) contains \ttt{274} jars released in the \ttt{org.apache.spark}, \ttt{org.apache.hadoop}, \ttt{org.apache.livy}, and \ttt{org.scala-lang} groups and their dependencies.

\paratitle{Spark and HDFS configurations.} In our setup, the HDFS data directories, such as dfs.datanode.data.dir, dfs.namenode.name.dir, hadoop.tmp.dir are pointed to the directories in the mounted block device. For simplicity, we keep the replication factor \ttt{1}. In this setup, we need on average \ttt{1 min 53 sec} to copy a single file of size \ttt{1GB} from local disk to HDFS with \ttt{hadoop fs -copyFromLocal} command. In addition, we also configured memory and virtual cores for Yarn and Spark-based on the number of nodes in the cluster and per node available resources. We defer the detailed discussion to Appendix~\ref{sec:memory_calculation}.

\begin{figure*}[t]
  \centering
  \subfigure[HiBench Large Profile]{
      \includegraphics[width=.3\textwidth]{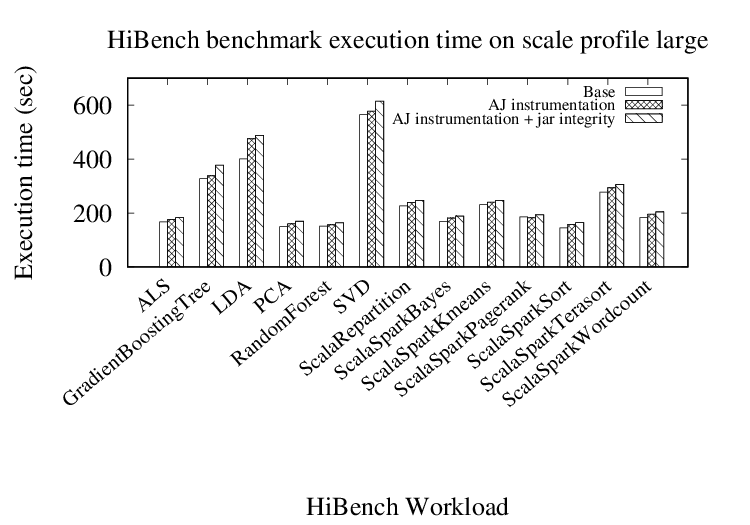}
      \label{fig:hibench_all_large}
  }
  \subfigure[Bayes on different scale]{
      \includegraphics[width=.30\textwidth]{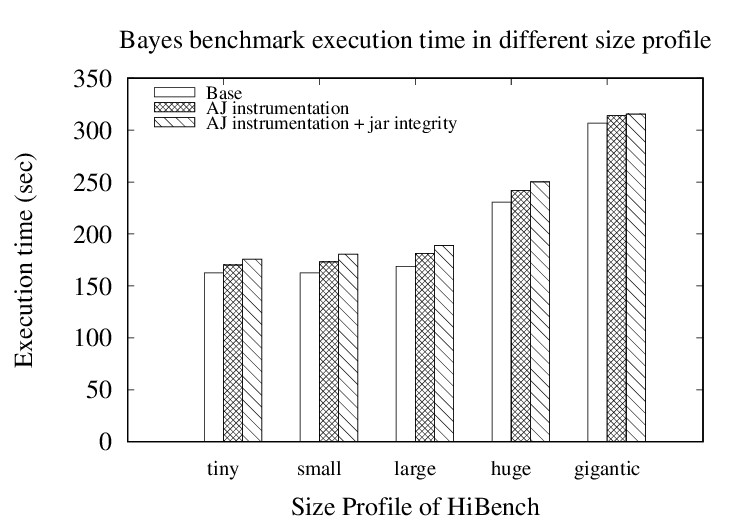}
      \label{fig:hibench_bayes}
  }
  \subfigure[RandomForest tree on different scale]{
      \includegraphics[width=.30\textwidth]{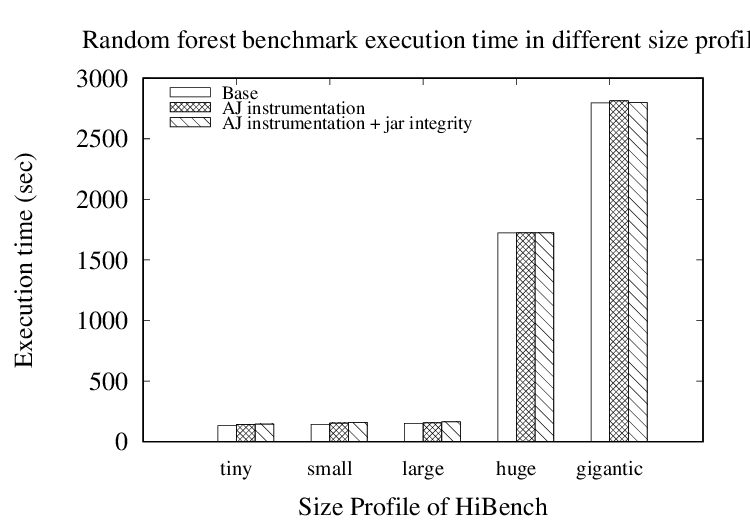}
      \label{fig:hibench_rf}
  }
  \vspace{-15pt}
  \caption{HiBench performance without any enforcement, with java agent enabled, and with integrity checking}
  \Description{HiBench performance without any enforcement, with java agent enabled, and with integrity checking}
  \label{fig:hibench}
\end{figure*}

\subsection{Performance of static components}
In this section, we present the performance analysis of the static components of our system that do not depend on the dataset or computation. Our evaluation answers the following questions.

\begin{itemize}
    \item What is the overall overhead of the static components?
    \item What is the accuracy of proactive defense? Does it block legitimate cases?
\end{itemize}

\paratitle{Performance overhead of static components.} Our proactive defense, jar rewriting, runtime jar instrumentation with AspectJ, and jar integrity checking are the static components of the system.
Since the proactive screening and jar rewriting can be done offline prior to running the data analytics tasks, here we report the performance overhead of Jar instrumentation with AspectJ to load access control policies and integrity checking for the TCB. We used HiBench~\cite{HiBench} benchmarks to see the performance overhead of these components on standard workloads. In Figure~\autoref{fig:hibench_all_large} we show the overall execution time of \ttt{13} SparkBench workload. We run the experiments on a Hadoop/Spark cluster with 5 nodes (1 master and 4 workers) nodes.
We observe that AspectJ instrumentations have a median overhead of \ttt{4.84\%} with Q1 \ttt{2.9\%} and Q3 \ttt{6.71\%} compared to the base case.
Similarly, we observe that jar integrity checking has median overhead  \ttt{4.28\%} with Q1 \ttt{4.06\%} and Q3 \ttt{6.00\%} compared to the AspectJ instrumentation case.
Finally, we examined the overhead of varying the input data size on two workloads. We used different scale profiles defined in HiBench to generate inputs of varying sizes. Specifically, we use \ttt{tiny}, \ttt{small}, \ttt{large}, \ttt{huge}, and \ttt{gigantic} profiles on \ttt{SparkScalaBayes} (in \autoref{fig:hibench_bayes}) and \ttt{RandomForest} (in \autoref{fig:hibench_rf}) workloads. For Bayes workload, we see linear growth over the size profiles with AspectJ overhead averaging \ttt{5.15\%} for AspectJ instrumentation, and with jar integrity checking we observe additional \ttt{3.12\%} overhead. In contrast, for random forest workload, we observe exponential growth in execution time with overhead averaging \ttt{3.55\%} for AspectJ instrumentation and \ttt{2.06\%} additional overhead for jar integrity checking. In summary, we observe an almost constant overhead in instrumenting and jar integrity checking, which was expected.

\begin{figure*}[t!]
    \centering
    \subfigure[Query 2]{
        \includegraphics[width=.30\textwidth]{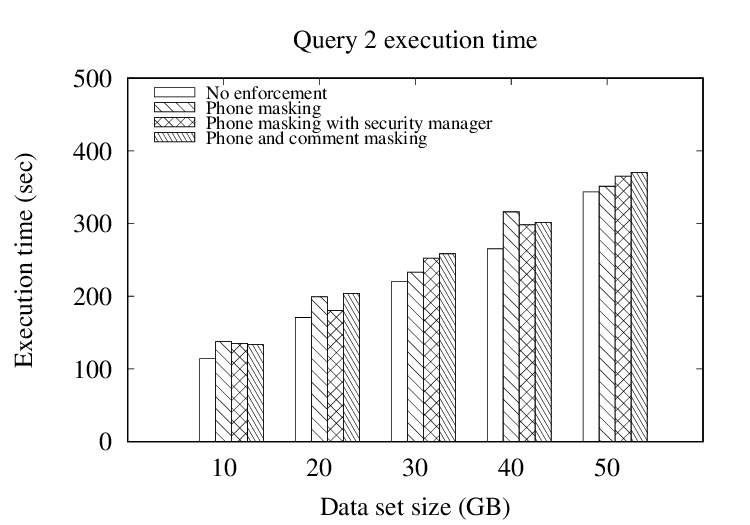}
        \label{fig:tpch_scale_q2}
    }
    \subfigure[Query 6]{
        \includegraphics[width=.30\textwidth]{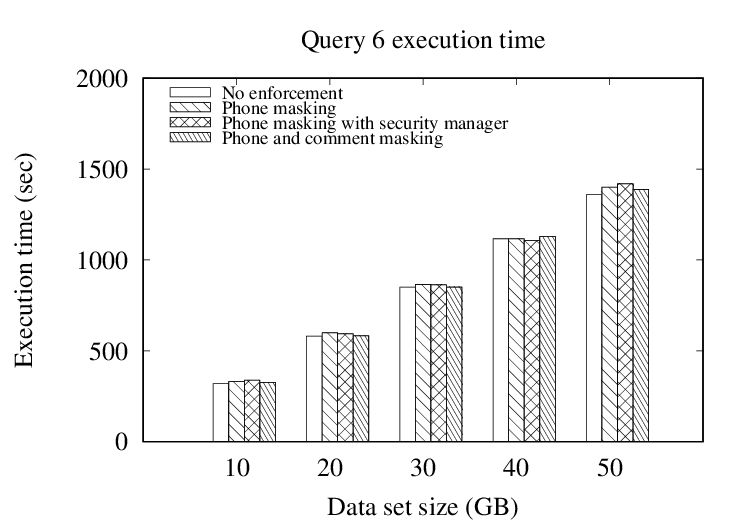}
        \label{fig:tpch_scale_q6}
    }
    \subfigure[Query 14]{
        \includegraphics[width=.30\textwidth]{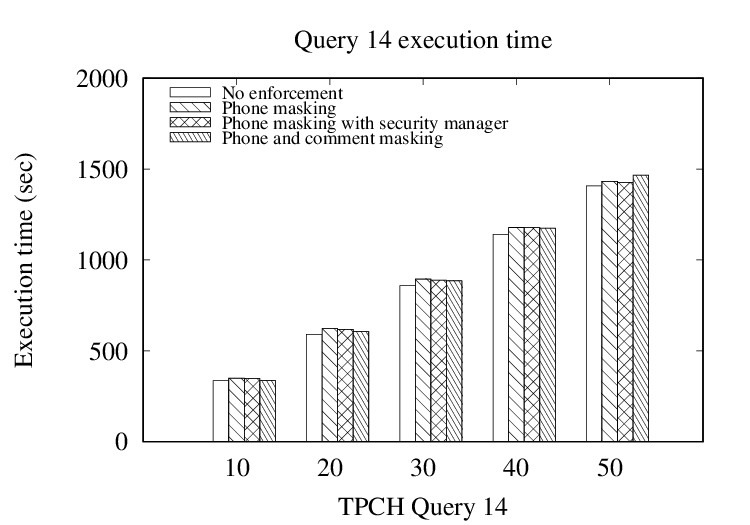}
        \label{fig:tpch_scale_q14}
    }
    \vspace{-15pt}
    \caption{Overhead of access control and security manager for TPCH queries on different input size.}
    \Description{Overhead of access control and security manager for TPCH queries on different input size.}
    \label{fig:tpch_vs_size}
\end{figure*}

\noindent
\textbf{Accuracy of proactive defense.} To check the soundness and precision of our proactive analysis on real-world code, we collected \ttt{\evalTotalRepos} repositories from GitHub that use Apache Spark. We specifically searched with keywords \evalKeywords{} We found \evalTotalMaven repositories use maven~\cite{maven} as a build tool. We focused on projects utilizing maven because in maven, the build script (\texttt{pom.xml}) is written in XML following a \textit{predefined} schema, which makes it easier to parse the project structure and find expected output binaries. Among \evalTotalMaven repositories, we successfully built \evalTotalBuiltSuccess~(\evalTotalBuiltSuccessPercent). 
From these repositories, we found \evalTotalAnalyzableJars analyzable jar files. We excluded jar files that contain dependencies, i.e., uber-jar~\cite{uber_jar}, also known as the fat jar. Because, during the uber jar generation process using maven shade plugin~\cite{shade_maven}, some class binaries are changed. So, an allow listing strategy based on class binary hash will not work in such a specialized scenario. The programmer needs to build hashes of custom class binaries to make our proactive analysis work in this scenario.
In addition, we also excluded some repositories (2) that extended the Apache Spark framework by copying a large portion of the original code and rebuilt core jars, which is highly unusual.
Among the analyzable jars, we found one or more issues in \evalJarWithIssues jars. In \evalRuleOrgApacheSparkPackage jars, our proactive analysis found attempts to define classes in \ttt{org.apache.spark} package. In these cases, programmers defined some classes under \ttt{org.apache.spark}, such as copying examples from \ttt{org.apache.spark.examples} package.
In \evalRuleInvokingRestrictedClasseForName jars, we observed invocation of \ttt{Class.forName}. In most cases, programmers load drivers of different database servers, such as MySQL and PostgreSQL, which were used to load data from/to Apache Spark. Also, in \evalRuleInvokingRestrictedNetwork jars, we have observed network class access, such as \ttt{java.net.Socket}, \ttt{java.net.URL}. In this case, programmers are trying to connect to a different host for downloading or uploading data. \textit{We conservatively blocked APIs for network access since this can be leveraged to temper with our TCB to step out of our threat model. Note that our allowlisting service can be used to exclude any legitimate uses from blocking.}
Among cases where the proactive analyzer failed to analyze the built jar, the most common reason is the internal error of the soot framework~\cite{soot}, which we used for building the analyzer. Using an updated soot library can potentially help with these failing cases.


\subsection{Overheads of dynamic components}
Our access control mechanism and reactive defenses are enforced at runtime, which depends on the dataset and the nature of the computation. In this section, we perform several experiments to extensively evaluate the overhead associated with them. Our experimental evaluation answers the following research questions.

\begin{itemize}
    \item How does the overhead of our attribute-based access control and reactive defense change over the baseline with the size of the dataset?
    \item What is the impact of the number of computing nodes?
\end{itemize}

\paratitle{Experimental setup.}
For this experiment, we use TPCH benchmark\footnote{http://www.tpc.org/tpch/}. We run TPCH queries on CSV data using Spark. More specifically, we store the TPCH \ttt{tbl} tables in HDFS as CSV files, load them as dataframes in Spark, and run the TPCH queries on them.
For these experiments, we set up two sets of policies
\begin{enumerate}[leftmargin=*,topsep=0pt]

    \item \emph{Masking on phone columns.} We show the last \ttt{4} digits of \ttt{12} digits on the \ttt{phone} column of \ttt{customer} and \ttt{supplier} table of TPCH.

    \item \emph{Masking on comments columns.} We use regular expressions to detect phone numbers and email address inside the comments column of \emph{all} the tables of TPCH and replace with defined patterns. We use regular expression \texttt{\textbackslash(?}\texttt{\textbackslash d{3}\textbackslash)?(-| )\textbackslash d{3}-\textbackslash d{4}} to detect phone numbers and replace with \ttt{`***-***-dddd'} pattern, where \ttt{d} represents a digit in the input string. This masking essentially shows only last 4 digits of the phone number. Similarly for email addresses, we use the regular expression 
    \texttt{\textbackslash b[\^ \ \textbackslash s]+@[a-zA-Z0-9][a-zA-Z0-9-\_]\{0,61\}[a-zA-Z0-9]\{0,1\} \textbackslash .([a-zA-Z]\{1,6\}|[a-zA-Z0-9-]\{1,30\}\textbackslash .[a-zA-Z]\{2,3\})\textbackslash b} to mask emails in the form of \ttt{*@*c}, to only show the last character of the email address. 
\end{enumerate}
Specific details of the policies are listed in \autoref{sec:policies}.

\paratitle{Impact of the dataset size on overheads.}
In this experiment, we generate and load \ttt{10}GB, \ttt{20}GB, \ttt{30}GB, \ttt{40}GB, and \ttt{50}GB of TPCH data in HDFS and run queries \ttt{2}, \ttt{6}, and \ttt{14} on them. To measure the runtime overhead, we used the following four settings, i) without any access control policy enforcement, ii) with phone masking policies, iii) with phone and comment masking policies, iv) with phone masking and security manager enabled. 

\begin{figure*}[t!]
    \centering
    \subfigure[Query 2]{
        \includegraphics[width=.30\textwidth]{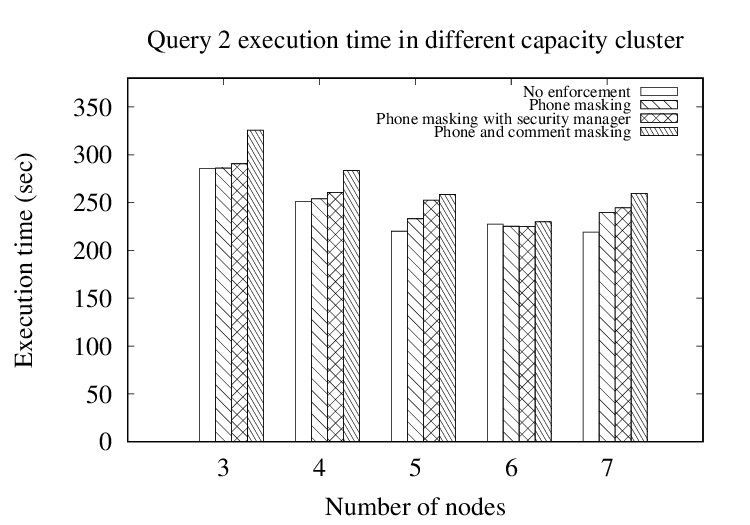}
        \label{fig:tpch_w_q2}
    }
    \subfigure[Query 6]{
        \includegraphics[width=.30\textwidth]{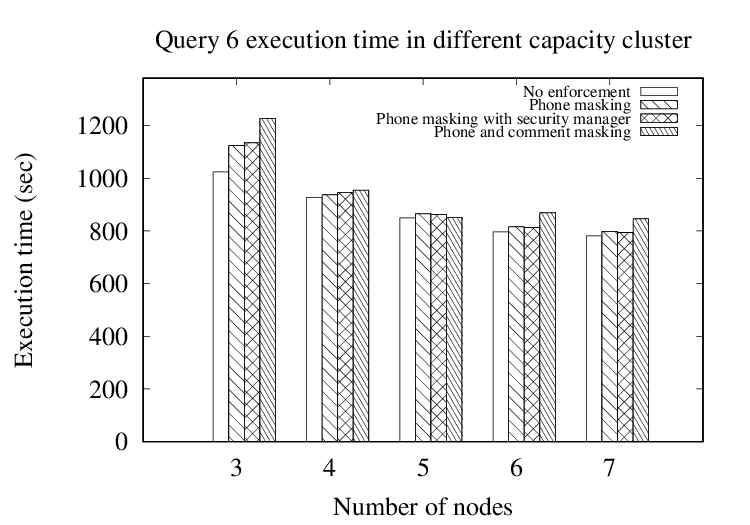}
        \label{fig:tpch_w_q6}
    }
    \subfigure[Query 14]{
        \includegraphics[width=.30\textwidth]{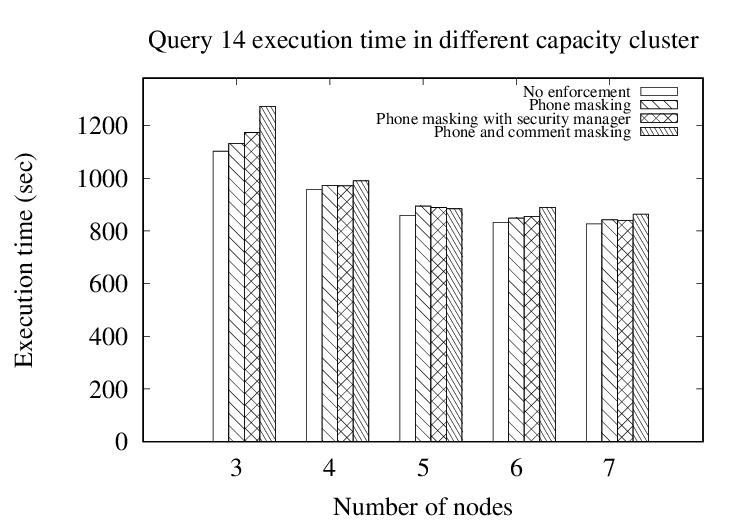}
        \label{fig:tpch_w_q14}
    }
        \caption{Overhead of access control and security manager for TPCH queries in different sized clusters}
    \Description{Overhead of access control and security manager for TPCH queries in different sized clusters.}
    \label{fig:tpch_vs_workers}

\end{figure*}

In \autoref{fig:tpch_vs_size} we illustrate the execution time of these queries. For query \ttt{2} (Figure~\autoref{fig:tpch_scale_q2}) we observe average overhead of \ttt{12.85\%} with standard deviation \ttt{8.17\%} for phone number masking. With security manager enabled we observe mean overhead of \ttt{11.44\%} with standard deviation \ttt{5.43\%}. 
Finally, phone and comment masking together, we observe similar average overhead \ttt{15.03\%} with standard deviation \ttt{4.48\%}. For query \ttt{2}, apart from few extreme cases, we observe overhead around \ttt{16\%}.
Now, for query \ttt{6} we observe mean overhead of \ttt{2.21\%} with standard deviation \ttt{1.42\%} for phone masking. With security manager enabled we observe mean overhead of \ttt{2.54\%}  with standard deviation \ttt{2.56\%}. Finally with phone and comment masking we observe mean overhead of \ttt{1.15\%} with standard deviation \ttt{0.94\%}. For query \ttt{6} our overhead is in and around \ttt{2\%}. 
For query \ttt{14}, we observe average overhead of \ttt{3.72\%} with standard deviation \ttt{1.4\%} with phone masking. With security manager we observe average overhead \ttt{3.33\%} with standard deviation \ttt{1.24\%}. Finally with phone and comment masking we observe mean overhead \ttt{2.54\%} with standard deviation \ttt{1.55\%}. Similar to query \ttt{6} we observe overall overheads of around \ttt{2\%} in query \ttt{2}.
To summarize, we do not observe any variation in runtime overhead over the baseline with the increase of the dataset size. However, the overhead is highly dependent on the type of query and the policies we enforce. In our case, the query \ttt{2} has complex inner query, several \textit{joins}, and \textit{order by} clauses, resulting noticeable overheads. In contrast, query \ttt{6} and \ttt{14} has complex aggregation and simpler joins, so the overheads are significantly lower.

\paratitle{Impact of computing nodes on overheads.}
To observe the overhead in varying computation capacity, we create a cluster with varying number of worker nodes, then load \ttt{30GB} of TPCH data, and run queries \ttt{2}, \ttt{6}, and \ttt{14}. We created cluster with \ttt{3}, \ttt{4}, \ttt{5}, \ttt{6}, and \ttt{7} Hadoop/Spark nodes and used one node as master and remaining as workers. We utilize equations outlined in \autoref{tab:yarn_spark_calculation} to calculate executor resource-related configurations.
For query \ttt{6} we have around \ttt{9.76\%} overhead with phone masking, which gradually decrease to about \ttt{1.77\%} for \ttt{5} nodes cluster and again increase a little bit in \ttt{7} nodes cluster. We observe a similar pattern in query \ttt{14}. Our hypothesis is increased computation capacity increases the parallelization in the cluster, hence, the overhead of our map and filter execution decreases. At some point, the overhead of parallelism, i.e. network and io overhead of exchanging data among nodes, will diminish this computational overhead. Furthermore, a very important point to emphasize is that Spark always greedily allocates \emph{all available memory}, whether it is needed or not. For query \ttt{2} on phone masking, we observe that overhead increases \ttt{1\%} to \ttt{9\%} for phone masking with increasing capacity. We observe a similar pattern for the combination of phone and comment masking as well.

\section{Discussion}

The soundness and accuracy of proactive analysis have significant security and usefulness implications. In theory, restriction (and false positives) comes with a price in usability. However, considering the difficulty of abuse detection, we believe this is a reasonable tradeoff.
Many real-world systems (e.g., Lua sandbox~\cite{luasandbox} and WeChat mini-apps~\cite{DBLP:conf/uss/WangZL23}) that allow execution of non-trusted code rely on API restrictions. 
Our evaluation on \evalTotalAnalyzableJars real-world jars indicates a low possibility of blocking legitimate uses. However, if our system raises a false alert for a jar or class file, a system admin can allowlist it and exclude it from analysis to ensure an uninterrupted operation. Our manual analysis of real-world projects did not discover any new suspicious features that our analysis might have missed. \textit{To further evaluate the detection capabilities, we created a set of 15 attacks, including all the attack scenarios discussed in} Section~\ref{sec:attack:irm:spark}. The evaluation shows that our proactive analysis \textit{successfully detected all the cases}. Guarding against the adversarial use of reflection APIs through static dataflow analysis is the unsound part of our system. Our jar re-write-based runtime checks theoretically guarantee their prevention.

\section{Related work}
\balance

\noindent
\textbf{Access control methodologies for data management systems.} 
Access control methodologies have been applied to data management systems over the years ranging from relational databases (e.g., see discussion and references in~\cite{access-control-db}) to non-relational systems (e.g.,~\cite{DBLP:conf/ccs/UlusoyCFKP15, ravi-att-hadoop, derbeko2016security,awaysheh2020next, sparkx, smartguard}). Compared to all these works, \securedl has the only framework-agnostic fine-grained attribute-based access control framework with data masking and filtering, which can support policy specification with Scala code snippets to handle any type of structured/relational data.

Another relevant data access control mechanism is purpose-aware access control (PAAC), where we define and enforce access control policies based on the purpose of the computation. PAAC and ABAC are complementary to each other. 
GuardSpark++~\cite{guardsparkplus} implements PAAC in Spark SQL for structured data (our access control framework works on both structured and non-structured data). It utilizes Spark's internal SQL query optimization engine `Catalyst' and enforces purpose-aware policies between analysis and optimization states.

\noindent
\textbf{Commercial solutions.} There are several commercial open-source projects (i.e., Apache Ranger) that offer access control atop distributed data analytics platforms. Intrinsically, the capability of Ranger is limited by the capability granted by the plugin system of the host framework. Since Apache Spark does not have a fine-grained access control plugin system, Ranger can not support it directly. All large cloud vendors have their own version of data analytics and access control mechanisms. For example, an Amazon Web Services customer can load CSV data in an s3 object store and run a HiveQL query using a service named Athena~\cite{aws_athena}. Access control in this scenario will be equivalent to the access control settings in the underlying data store s3. These solutions revolve around solving the access control on structured data.

\noindent
\textbf{Static code analysis for vulnerability detection.} Static code analysis has been extensively used to detect API misuse vulnerabilities Java code~\cite{DBLP:conf/ccs/EgeleBFK13, rahaman2019cryptoguard, DBLP:conf/ecoop/KrugerS0BM18, DBLP:conf/ccs/FahlHMSBF12, DBLP:conf/ccs/BosuLYW17, oltrogge2021eve, DBLP:conf/ndss/NanY0ZZ018, DBLP:conf/ndss/BianchiFMKVCL18, zuodoes, DBLP:conf/uss/MahmudAAER20}. Most of the work focuses on detecting system-level API misuses~\cite{DBLP:conf/ccs/EgeleBFK13, rahaman2019cryptoguard, DBLP:conf/ecoop/KrugerS0BM18, DBLP:conf/ccs/FahlHMSBF12, DBLP:conf/ndss/NanY0ZZ018, DBLP:conf/ndss/BianchiFMKVCL18}, such as SSL/TLS~\cite{rahaman2019cryptoguard, DBLP:conf/ccs/FahlHMSBF12}, Cryptographic APIs~\cite{DBLP:conf/ccs/EgeleBFK13, rahaman2019cryptoguard, DBLP:conf/ecoop/KrugerS0BM18}, APIs for fingerprint protection~\cite{DBLP:conf/ndss/BianchiFMKVCL18}, Android Inter-app communication APIs~\cite{DBLP:conf/ccs/BosuLYW17}, etc. Some of the recent works focus on non-system APIs too~\cite{zuodoes, DBLP:conf/uss/MahmudAAER20}, such as  cloud service APIs for information storage~\cite{zuodoes}, Creditcard information processing APIs~\cite{DBLP:conf/uss/MahmudAAER20}, etc. In this scenario, no missed detection is expected-but-not-critical. In our case, a missed detection has a serious consequence on the overall security guarantee. Consequently, we employ runtime checks to detect and block such cases.

\section{Conclusion}
Typically, fine-grained access controls are enforced using aspect-oriented programming on top distributed data analytics platforms that do not have any plugin support. In this work, we show that it is possible to evade such access-checking mechanisms without leaving traces, which we call transient evasion attacks. Next, we designed a two-layered defense to enable secure enforcement under such attacks. We are the first to utilize the program analysis to complement the existing security features and use code rewriting to design the defense mechanism. We also propose a new framework agnostic fine-grained access control framework with enhanced policy language. Finally, we leveraged our defense mechanism to securely implement the proposed access control on top of Apache Spark (which we named \securedl). We show \securedl's effectiveness with a prototype implementation. Our extensive experimental evaluation shows that the \securedl has a low overhead while securely enforcing attribute-based access control policies.

\begin{acks}
This research is supported by NFS SBIR Phase I 1647681, Phase II 1758628, and NIH SBIR Phase I 1R43LM013960-01 grants. Additionally, Sazzadur Rahaman was supported by the startup package from the University of Arizona. We also extend our gratitude to all software engineers of \href{https://www.datasectech.com}{DataSecTech}, Tasneem Yasmin, Sourav Dasgupta, Shakti Shivaputra, Xiaowen Ding, Praneeth Khanna Bala, Falak Singhal, Sayeed Salam, and Sudhanva Purushotham, for their valuable technical contributions in all DataSecTech projects and their collaborative spirit. Finally, we thank Lei Cao for his insightful feedback on an earlier draft of the paper.
\end{acks}

\clearpage

\balance
\bibliographystyle{ACM-Reference-Format}

\appendix

\section{Attack on GuardMR}
\label{sec:attack:guradmr}

Following is the code snippet that evades GuardMR protection on Apache Hadoop.

\begin{lstlisting}[caption={Reading file splits directly with a custom RecordReader in Hadoop to sidestep GuardMR.},label={lst:guardmr_attack}]
class MalReader extends RecordReader {

public void initialize (...) {
 List<FileSplit> splits = (List<FileSplit>) fileInputFormat.getSplits(job);
    
 for(FileSplit split : splits) {
    
  final FutureDataInputStreamBuilder builder =
      file.getFileSystem(job)
      .openFile(split.getPath());
    
  FSDataInputStream fileIn = FutureIOSupport
      .awaitFuture(builder.build());
    
  long start = split.getStart();
  long end = start + split.getLength();
    
  int length = 1024 * 1024;
  byte[] buffer = new byte[length];
  long position = start;
    
  while (position < end) {
    position += fileIn.readBytes(position, buffer, 0, length)
    // Access the plain text values
  }
 }
}
\end{lstlisting}

\section{Spark AOP details}
\label{sec:spark_aop_details}

We create \emph{pointcut} for all the user-facing methods in Spark that are used to read files and create RDD or DataFrame, such as
\ttt{org.apache.spark.SparkContext.textFile(String, int)}, \ttt{org.apache.spark.sql.DataFrameReader.json(String)}, etc. An Spark data user can utilize these methods to create an initial RDD and then perform various operations. We find these methods by searching statements that instantiate RDD and its subclasses in the Spark source code. We then look for public methods that use these methods.
We attach AspectJ javaagent to all Spark's Java processes by modifying the shell script, \ttt{spark-class}. In addition, to communicate with other services, such as policy service, we need to initialize some service clients in spark context. So, we implement and attach a \ttt{SparkListenter}. Specifically, we modify these five configurations -
\ttt{spark.driver.extraJavaOptions}, \ttt{spark.executor.extraJavaOptions}, 

\noindent
\ttt{spark.driver.extraClassPath},

\noindent
\ttt{spark.executor.extraClassPath},
\ttt{spark.extraListeners}.

We use \texttt{@Around()} and related annotations from AspectJ to attach policy enforcement code to data reading methods. One such example is listed in     
 Listing~\ref{lst:policy_enforcement_around}. Here we define a method \ttt{policisOnTextFile} with an \ttt{@Around("execution(* org.apache.spark.SparkContext.textFile(String, int))")} annotation, which signals AspectJ to attach the \ttt{policisOnTextFile} method around \ttt{SparkContext.textFile} method. In other words, any time the spark context method is executed we first receive the call in our \ttt{policisOnTextFile} method. This method take as argument \ttt{ProceedingJoinPoint} class and return either a modified RDD with access control enforcement or raise error due to insufficient access permission. 
We get all the original input parameters, such as the file path, using \ttt{getArgs} method of the joint point class. Next, we decide whether the user has access to the file using file metadata information. If the user has access to the file, we execute the \ttt{proceed} on the joint point. This creates the initial RDD or DataFrame. To emphasize, this does not actually execute the read access; instead, it creates a job in a DAG that will be executed later. Now, we modify this RDD or DataFrame with access control policy implementation. 

\begin{lstlisting}[caption={\securedl advice with point cut using AspectJ annotation},label={lst:policy_enforcement_around}]
@Around("execution(* org.apache.spark
  .SparkContext.textFile(String,int))")
def policisOnTextFile(joinPoint):
    
    file_path <- joinPoint.getArgs[0]
    u <- fetch_user_info()
    if (!hasAccess(u, file_path)) {
        throw new AccessControlException()
    }
    
    rdd <- joinPoint.proceed()
    return enforce_policies(file_path, rdd)
\end{lstlisting}

In the policy enforcement method, we fetch policies and user information from our central policy server. Then, we collect connection information, such as the user's IP address. Next, we serialize and distribute the policies. In particular, we create executable byte code of the filters and masks in the matching policies and distribute the executables by using a central distributed cache server. Finally, we attach a \ttt{filter} and a \ttt{map} method with the input DataFrame or RDD. In the filter method, we execute the serialized filter method from the matching policies, and in the map method, we execute data masking policies. In Listing~\ref{lst:policy_enforcement}, we outline the implementation of the policy enforcement method.

\begin{lstlisting}[caption={\securedl policy enforcement implementation},label={lst:policy_enforcement}]

def enforce_policies(file_path, rdd):
    p <- fetch_policies(file_path)
    u <- fetch_user_info()
    c <- connection_info()
    f, m <- serialize_and_distribute(p)
    rdd.filter(t -> f(t, u, c) )
        .map(t -> m(t, u, c)

    return rdd
\end{lstlisting}

The example code of Listing~\ref{rdd:example} with policy enforcement will have two more modification methods, a filter, and a map, just after reading the file as listed in 
 Listing~\ref{rdd:example_with_enforcement}.
Although we did not list explicitly here, we implemented similar enforcement for all available DataFrame and RDD creation methods.
To summarize, we attach access control enforcement policies using APO. We attach our advice to spark's data reading methods and ensure these get executed by modifying the appropriate spark parameters in the spark execution script.

\begin{lstlisting}[caption={An example of applying access control before executing any user defined transformations},label={rdd:example_with_enforcement}]
long count = sc.textFile("users.csv")
   <@{\noindent
   \fbox{%
    \parbox{2in}{%
        \textcolor{red}{.filter(t -> f(t, u, c))} \\
        \textcolor{red}{.map(t -> m(t, u, c))}
    }%
}}@>  
   .map(line -> line.split(";"))
   .map(fields -> 
        Integer.parseInteger(fields[1]))
   .filter(salary -> salary > 100000)
   .count()

\end{lstlisting}

\paratitle{Implementation completeness.} One of the biggest challenges in our implementation is ensuring that we are trapping all methods. Otherwise, an attacker can bypass the security mechanism by reading data using those methods. Therefore, to complete our implementation, we examined all available official tutorials and thoroughly went over the source codes of related packages in Spark for the listed methods of reading data in Spark. Furthermore, if new data reading methods are introduced later, we can easily write `advices' for these methods. However, we observe that data reading method changes are infrequent. Apache Spark tends to keep the user-facing API consistent over version updates. Therefore, machine learning models written in one version can run on a different version without further modification.

\section{Implementation with Hive Plugins}
\label{hive:implimentation}

We trivially implement our access control framework in Hive by leveraging Hive's plugin system.
Specifically, we wrote an authorizer class by extending the public interface \ttt{org.apache.hadoop.hive.*.plugin.HiveAuthorizer} to integrate our access control checking logic. In addition, we implemented a factory class extending \ttt{org.apache.hadoop.hive.*.plugin.HiveAuthorizerFactory} interface. In this class, we instantiate the authorizer class with proper parameters. Finally, to configure the hive authorization process properly, we set configuration variable \ttt{hive.security.authorization.enabled} to \ttt{true} and \ttt{hive.security.authorization.manager} to the full classpath of our authorizer class.

\begin{figure}[h!]
\begin{center}
\includegraphics[width=.4\textwidth]{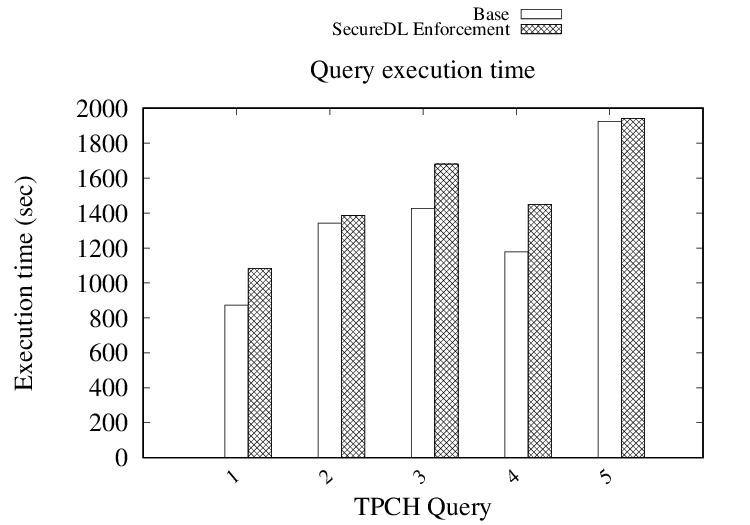}
\end{center}
\caption{Overhead of running different TPCH queries on 100GB data in Hive with our access control enabled.}
\label{fig:tpch_hive_100g}
\end{figure}

To test the overhead of our hive reactive enforcement, we load \ttt{100GB} of TPCH data on Hive and execute TPCH queries \ttt{1} to \ttt{5}. In \autoref{fig:tpch_hive_100g}, we show the overheads. We observe that overheads range from \ttt{0.88\%} to \ttt{23.94\%}. This wide range in overhead is due to the fact that a few queries contain operations on policy-controlled columns and others do not.

\section{Memory Calculation}
\label{sec:memory_calculation}

We adopt the technique outlined in~\cite{cloudera_yarn_calculation} to allocate resources for the YARN container. We exclusively consider two resources - memory and virtual cores (vCores). We reserve some memory and vCores for system processes, then divide the remaining memory into containers. We use \textit{number of containers} $\containers$ and \textit{memory per container} $\containerram$ as input to calculate the total memory allocation available for node manager, minimum and maximum memory allocation limit, application manager allocation limit, map-reduce memory allocations, and vCores allocation limit.
In our setup, we calculate the number of vCore per node by multiplying \ttt{2.5} to available OCPU, since the OCI OCPU are not shared with other tenants~\cite{ocpu_vcpu_comparison}. So, for a \ttt{VM.Standard2.4} node calculates the number of available vCores is \ttt{10 = 4 $\times$ 2.5}. We reserve \ttt{2} vCores and \ttt{12GB} of memory for system processes. That leaves us with \ttt{8} vCores and \ttt{48GB} of memory for yarn container in a node. We set minimum container resources to \ttt{1} vCores and \ttt{6GB}. 
We use the equations outlined in \autoref{tab:yarn_spark_calculation} to calculate all relevant memory configurations for yarn. We also need to tune resources for Apache Spark. In particular, we need to divide the available vCores and memory into \textit{executors}.
Given $\nodecores$ vCores in a worker node,
$\executorcores$ vCores per executor, $\nodeexecutors$ executors per worker and $\workers$ workers in total, we can calculate the number of executors per node as $\nodeexecutors = \lfloor \frac{\nodecores}{\executorcores} \rfloor $. Multiplying this value with the number of worker nodes gives us the total available executors. We reserve one executor for resource negotiation. For memory allocation per executor, we divide the total available memory per node by the number of executors per node. In our setup, we decided to go with \ttt{2} vCores per executor setup. So, on a \ttt{4} worker nodes cluster the relevant parameters are \ttt{--executor-cores 2} \ttt{--num-executors 15}, and \ttt{--executor-memory 10752MB}.

\begin{table}[h]
\begin{tabular}{ |l|l| } 
 \hline
Configuration  & Equation \\
\hline
yarn.nodemanager.resource.memory-mb & = $\containers * \containerram$ \\
yarn.scheduler.minimum-allocation-mb    & = $\containerram$ \\
yarn.scheduler.maximum-allocation-mb    & = $\containers * \containerram$ \\
mapreduce.map.memory.mb & = $\containerram$ \\
mapreduce.reduce.memory.mb  & = $2 * \containerram$ \\
mapreduce.map.java.opts & = $0.8 * \containerram$ \\
mapreduce.reduce.java.opts  & = $0.8 * 2 * \containerram$ \\
yarn.app.mapreduce.am.resource.mb   & = $2 * \containerram$ \\
yarn.app.mapreduce.am.command-opts  & = $0.8 * 2 * \containerram$ \\
 \hline
Executors per node   & \textbf{}$ \nodeexecutors = \lfloor \frac{\nodecores}{\executorcores} \rfloor $ \\
Number of executors   & = $  \nodeexecutors * \workers - 1$ \\
Executor memory   & = $ \frac{\containers * \containerram}{\nodeexecutors}   $ \\
 \hline
\end{tabular}
\caption{Yarn and Spark resource calculation formulas}
\label{tab:yarn_spark_calculation}
\vspace{-1pt}
\end{table}

\section{Policies}
\label{sec:policies}

\begin{lstlisting}[caption=Policy example]
Masks:
    phone:
      name: PhoneNumberMask
      type: regex_mask
      detection_regex: "\\(?\\d{3}\\)?(-| )
      \\d{3}-\\d{4}"
      replacement_pattern: '***-***-dddd'

    email:
      name: EmailMask
      type: regex_mask
      data_type: email
      detection_regex: "\\b[^\\s]\
      +@[a-zA-Z0-9]\
      [a-zA-Z0-9-_]{0,61}\
      [a-zA-Z0-9]{0,1}\
      \\.([a-zA-Z]{1,6}|\
      [a-zA-Z0-9-]{1,30}\
      \\.[a-zA-Z]{2,3})\b"
      replacement_pattern: '*@*c'

    l4of12d:
      type: static_mask
      data_type: digit
      length: 12
      name: ShowLast4Of12Digits
      visible_anchor: end
      visible_chars: 4

Policy:
  customer_accounts:
    document: customers.accounts
    filter:   |
      val ip : String 
        = context("ip").asInstanceOf[String]
      val z : Integer
        = row("zip").asInstanceOf[Integer]

      if (ip == "10.5.17.19") {
          // Zeppelin IP

          z == 75080
      } else if(ip == "10.5.17.10") {
          // Command line IP

          z >= 75080 \&\& z <= 75081
      } else {
          false
      }

    masks:
      credit_card:
        - Masks.l4of12d
      comments:
        - Masks.email
        - Masks.phone
\end{lstlisting}

\end{document}